# Defect Engineering: Graphene Gets Designer Defects

Lincoln D. Carr and Mark T. Lusk
*Colorado School of Mines, Golden, Colorado, USA*
e-mail: lcarr@mines.edu; mlusk@mines.edu

Defects are integral to the semiconductor industry, with the addition of just one dopant atom per hundred million host atoms being able to significantly change the electronic properties of the host material. Each dopant can be seen as a kind of atomic point defect. Defects can also be patterned into extended structures to give entirely new properties. Theorists have proposed that such extended defects could be used to modify the electronic properties of materials, notably graphene.

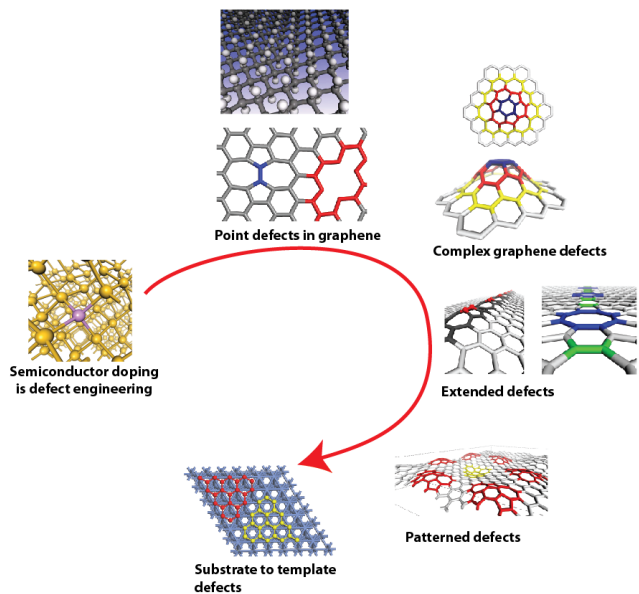

Graphene is an attractive target because its two-dimensional nature makes it easier to add, remove or move carbon atoms to alter its electronic properties, in a process known as self-doping[1-4].

Now, writing in *Nature Nanotechnology*, Matthias Batzill and colleagues[5] from the University of South Florida report the first experimental introduction of extended defects into graphene. Extended defects are built up from localized defects, which act as building blocks. The most common of these is the Stone–Thrower–Wales (STW) defect[6, 7], which results from a simple rotation of two carbon atoms, and produces two pentagons and two heptagons in the hexagonal graphene lattice (Fig. 1). Other building blocks are the inverse STW defect (with an inverse arrangement of pentagons and heptagons[2, 3, 8, 9] and the di-vacancy defect (which is formed by the removal of two adjacent carbon atoms10). The South Florida team introduced alternating STW and di-vacancy defects to create an extended defect made of paired pentagons and octagons in an epitaxial layer of graphene on a nickel substrate. Carbon atoms can be arranged on nickel in a variety of ways, and these arrangements have different energies. However, two of these arrangements (in which the carbon atoms rest in lattice 'hollows' that are either one or two nickel layers deep) have nearly the same energy (Fig. 1c), so small fluctuations in energy lead to scattered domains of each. At the interface between these domains, a dislocation forms, and this is the line defect. When the nickel support is chemically dissolved, a flake of graphene containing extended defects is left behind.

Although Batzill and colleagues did not demonstrate directly that their line defect was conducting, the density of states that they measured for the defect was similar to that of the metallic edge states of graphene flakes. This suggests that their defect could function as a conducting nanowire in graphene-based electronic devices. The defect was observed with a scanning tunnelling microscope, and interpreted with density-functional theory. Such combined use of experimental and theoretical tools is at the heart of current nanoengineering practice. In the current experiment, extended line defects were promoted by the nickel template but were not precisely arranged. An outstanding question is therefore: can such defects be introduced in a controlled manner? This would allow a variety of novel applications. Directed placement
in graphene could guide charge as well as spin (for spintronic devices), and atoms and molecules (for microfluidics).
Extended defects might also serve as sites for controlled chemical reactions or, as Batzill and colleagues mention, as porous membranes through which only very small molecules could pass. The membrane structure of graphene also allows for defect structures more complex than points or lines, which could lead to other useful electrical, magnetic and chemical properties. In addition, it should be possible to develop templated defects in a similar fashion in materials other than graphene.

We close by observing an analogy to string theory, which proposes that the fundamental building blocks of matter are one-dimensional strings (or higher dimensional objects called branes), rather than point-like particles. String theory is the leading candidate for a theory to unify all the fundamental forces of nature, and has influenced our understanding of everything from quantum mechanics to relativity. In the same fashion, by moving us from points to lines and even more complex objects, extended defect structures provide a new set of building blocks for nanomaterials that could lead to new physical effects and device concepts.

**References**
1. Crespi, V. H., Benedict, L. X., Cohen, M. L. & Louie, S. G. *Phys. Rev. B* **53,** 13303–13305 (1996).
2. Terrones, H., Terrones, M., Hernandez, E., Grobert, N., Charlier, J. C. & Ajayan, P. M. *Phys. Rev. Lett*. **84,** 1716–1719 (2000).
3. Lusk, M. T. & Carr, L. D. *Phys. Rev. Lett.* **100,** 175503 (2008).
4. Lusk, M. T. & Carr, L. D. *Carbon* **47,** 2226–2232 (2009).
5. Lahiri, J., Lin, Y., Bozkurt, P., Oleynik, I. I. & Batzill, M.
*Nature Nanotech.* **5,** 326–329 (2010).
6. Stone, A. & Wales, D. J. *Chem. Phys. Lett.* **128,** 501–503 (1986).
7. Thrower, P. A. in *Chemistry and Physics of Carbon* Vol. 5 (ed. Walker, Jr P. L.) 262 (Dekker, 1969).
8. Li, L., Reich, S. & Robertson, *J. Phys. Rev. B* **72,** 184109 (2005).
9. Orlikowski, D., Buongiorno-Nardelli, M., Bernholc, J. & Roland, C. *Phys. Rev. Lett.* **83,** 4132–4135 (1999).
10. Hashimoto, A., Suenaga, K., Gloter, A., Urita, K. & Iijima, S. *Nature* **430,** 870–873 (2004).